\documentclass[runningheads]{llncs}
\usepackage{amssymb}
\usepackage{graphicx}
\usepackage{cite}
\usepackage{amsmath,amssymb,amsfonts}
\usepackage{algorithmic}
\usepackage{textcomp}
\usepackage{xcolor}
\usepackage{subfig}
\usepackage{multirow}

\numberwithin{equation}{section}

\begin{document}
%
\title{Exploring the Role of Visual Content \\ in Fake News Detection}

%
%


\author{Juan Cao\inst{1,2} \and Peng Qi\inst{1,2}\orcidID{0000-0002-6458-8320} \and Qiang Sheng\inst{1,2}\orcidID{0000-0002-2481-5023} \and Tianyun Yang\inst{1,2} \and Junbo Guo\inst{1} \and Jintao Li\inst{1}}
\authorrunning{J Cao et al.}
%
\institute{
Key Laboratory of Intelligent Information Processing \& \\
Center for Advanced Computing Research, \\
Institute of Computing Technology, CAS, China
\email{\{caojuan,qipeng,shengqiang18z,yangtianyun19z,guojunbo,jtli\}@ict.ac.cn}\\
\and
University of Chinese Academy of Sciences, China
}



%
\maketitle              
\begin{abstract}
The increasing popularity of social media promotes the proliferation of fake news, which has caused significant negative societal effects. Therefore, fake news detection on social media has recently become an emerging research area of great concern. With the development of multimedia technology, fake news attempts to utilize multimedia content with images or videos to attract and mislead consumers for rapid dissemination, which makes visual content an important part of fake news. Despite the importance of visual content, our understanding about the role of visual content in fake news detection is still limited. This chapter presents a comprehensive review of the visual content in fake news, including the basic concepts, effective visual features, representative detection methods and challenging issues of multimedia fake news detection. This chapter can help readers to understand the role of visual content in fake news detection, and effectively utilize visual content to assist in detecting multimedia fake news.

\end{abstract}

\section{Introduction}
\subsection{Motivation}

Social media platforms, such as Twitter\footnote[1]{https://twitter.com/} and Chinese Sina Weibo\footnote[2]{https://weibo.com/}, have become important access where people acquire the latest news and express their opinions freely\footnote{http://www.cac.gov.cn/2019-08/30/c\_1124938750.htm}\footnote{https://www.journalism.org/2018/09/10/news-use-across-social-media-platforms-2018/}. However, the convenience and openness of social media have also promoted the proliferation of fake news, i.e., news with intentionally false information, which not only disturbed the cyberspace order but also caused many detrimental effects on real-world events. For example, in the political field, during the month before the 2016 U.S. presidential election campaign, the Americans encountered between one and three fake stories on average from known publishers\cite{fakenewsdefinition}, which inevitably misled the voters and influenced the election results; In the economic field, a piece of fake news claiming that Barack Obama was injured in an explosion wiped out \$130 billion in stock value\footnote{https://www.telegraph.co.uk/finance/markets/10013768/Bogus-AP-tweet-about-explosion-at-the-White-House-wipes-billions-off-US-markets.html}; In the social field, dozens of innocent people were beaten to death by locals in India because of a piece of fake news about child trafficking that was widely spread on social media\footnote{https://www.washingtonpost.com/world/asia\_pacific/as-mob-lynchings-fueled-by-whatsapp-sweep-india-authorities-struggle-to-combat-fake-news/2018/07/02/683a1578-7bba-11e8-ac4e-421ef7165923\_story.html}. Hence, the automatic detection of fake news has become an urgent problem of great concern in recent years \cite{surveykumar2018false, surveyKai2017Fake, surveyzubiaga2018}.

The development of multi-media technology promotes the evolution of self-media news from text-based posts to multimedia posts with images or videos, which attracts more attention from consumers and provides more credible storytelling. On the one hand, as a vivid description form, the visual content including images and videos is more attractive and salient than plain text and consequently boosts the news propagation. For instance, tweets with images get 18\% more clicks, 89\% more likes, and 150\% more retweets than those without images\footnote{https://www.invid-project.eu/tools-and-services/invid-verification-plugin/}. On the other hand, visual content is often used as evidence of a story in our common sense, which can increase the credibility of the news\footnote{https://www.businesswire.com/news/home/20190204005613/en/Visual-SearchWins-Text-Consumers\%E2\%80\%99-Trusted-Information}. Unfortunately, this advantage is also taken by fake news. For rapid dissemination, fake news usually contains misrepresented or even tampered images or videos to attract and mislead consumers. As a result, visual content has become an important part of fake news that cannot be neglected, making multimedia fake news detection a new challenge.

Multimedia fake news detection aims at effectively utilizing the information of several modalities, such as textual, visual and social modalities, to detect fake news. Visual modality can provide abundant visual information, which is preliminarily proven to be effective in fake news detection \cite{attRNN}. However, although the importance of exploiting visual content have been revealed, our understanding about the role of visual content in fake news detection remains limited. To further facilitate research on this problem, we present a comprehensive review of the visual content in fake news in this chapter, including the problem definition, available visual characteristics, representative detection approaches and challenging problems. 

\subsection{Problem Definition}
In this subsection, we introduce the concept of fake news and analyze the different types of visual content in fake news.

Fake news is widely defined as news articles that are intentionally and verifiably false and could mislead consumers \cite{fakenewsdefinition, thescienceoffakenews, surveyKai2017Fake}. On the context of social multimedia, news articles refer to news posts with multimedia content that are published by users, so the general definition of fake news has been further refined \cite{mediaeval15, mediaeval16, ijmir, fauxtography}. Formally, we state the refined definition as follows, \\

\noindent \textbf{Definition 1.1}  \textbf{A piece of fake news is a news post that shares multimedia content that does not faithfully represent the event that it refers to}.\\

\begin{figure}
	\vspace{-0.5cm}
	\setlength{\belowcaptionskip}{-0.5cm}
	\centering
	\subfloat[]{
		\label{Fig: manipualted}
		\includegraphics[width=0.3\textwidth]{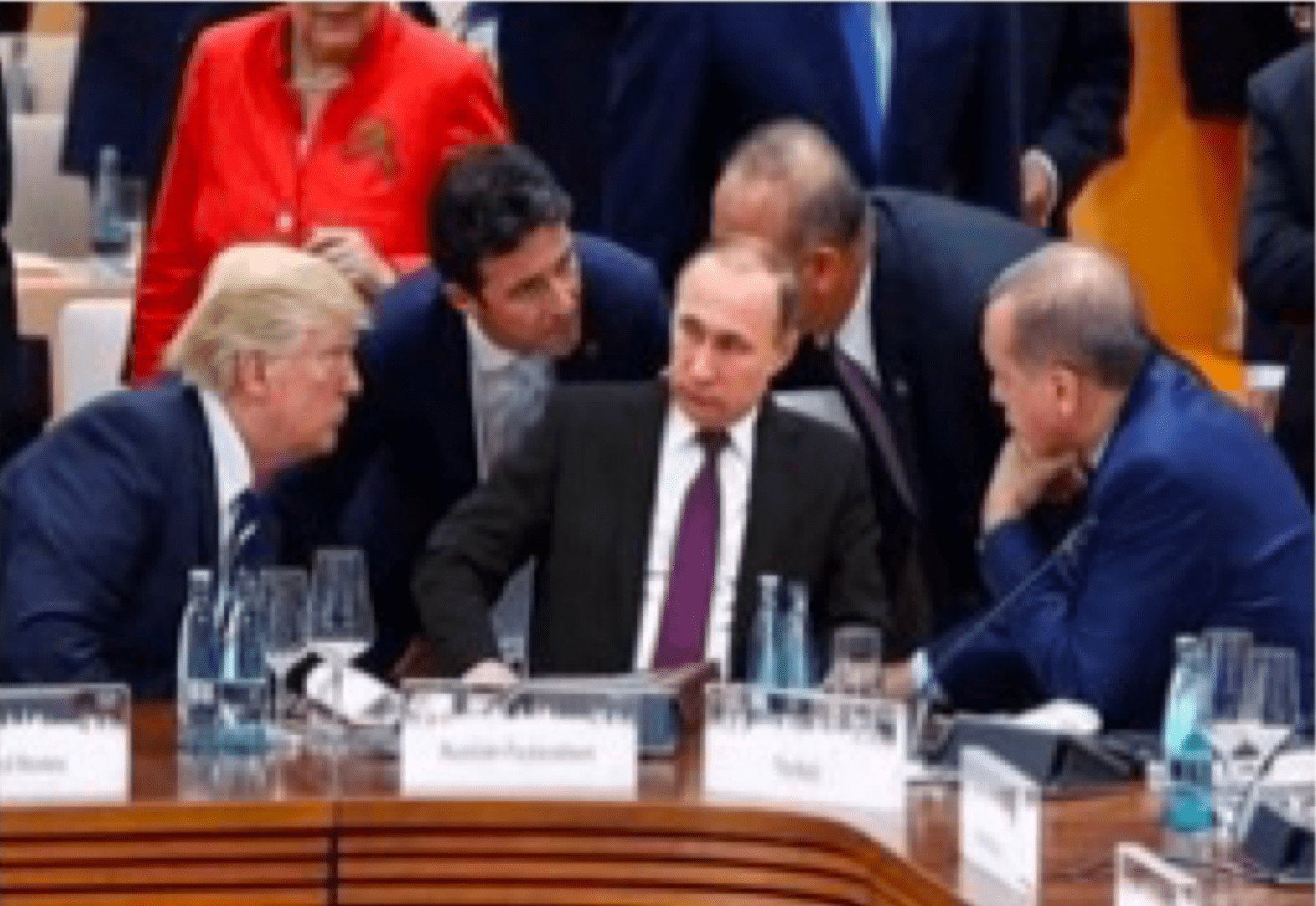}
	}
	\subfloat[]{
		\label{Fig: irrelevant}
		\includegraphics[width=0.3\textwidth]{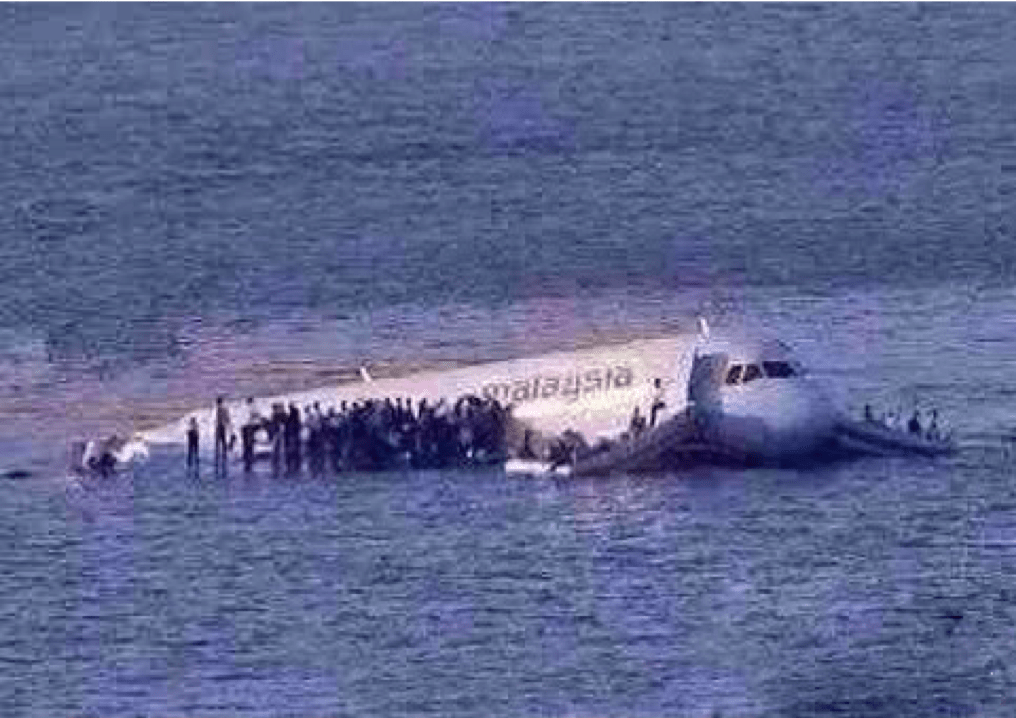}
	}
	\subfloat[]{
		\label{Fig: misrepresent}
		\includegraphics[width=0.3\textwidth]{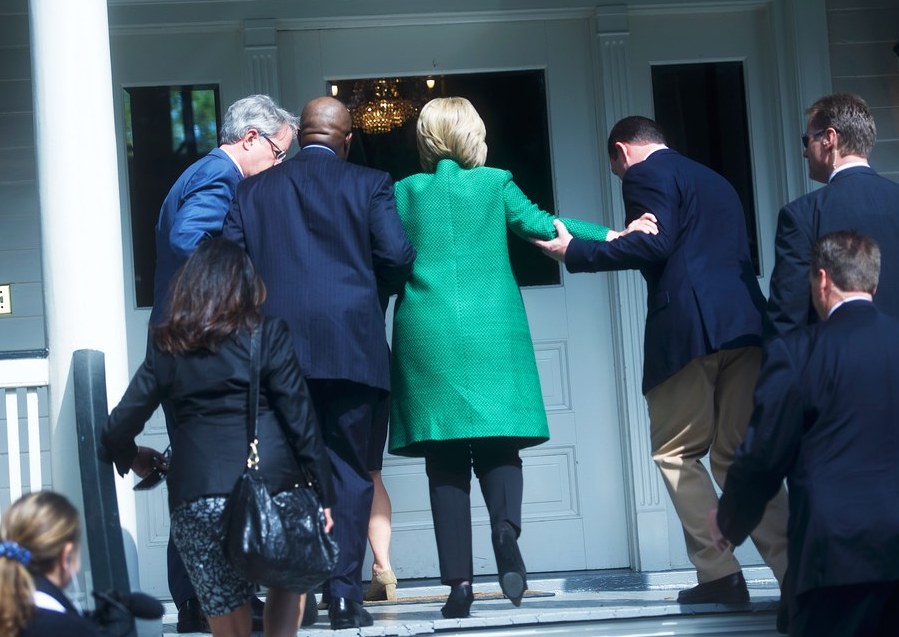}
	}\\
	\caption{Examples of the visual content in fake news: (a) A tampered image where Putin is spliced on the middle seat at G-20 to show that he is in the center position of an intense discussion among other world leaders; (b) A real image captured in 2009 New York air crash, but it is claimed to be the wrecked Malaysia Airlines MH370 in 2014; (c) A real image taken at the moment when Hillary Clinton accidentally stumbled, but it was maliciously interpreted as evidence of Clinton's failing health.}
	\label{Fig: cases}
\end{figure}

In real-world scenarios, the visual content in fake news can be broadly classified into three categories: (1) visual content that is deliberately manipulated (also known as tampering, doctoring or photoshopping) or automatically generated by deep generative networks, which equals to \textit{fake images/videos} in our common sense (see Figure \ref{Fig: manipualted}), (2) visual content from an irrelevant event, such as a past event, a staged work or an artwork, that is reposted as being captured in the context of an emerging event (see Figure \ref{Fig: irrelevant}), or (3) visual content that is real (not edited) but is published together with a false claim about the depicted event (see Figure \ref{Fig: misrepresent}). All examples in Figure \ref{Fig: cases} fall under our definition of fake news, because the images and associated texts jointly convey the misleading information regardless of the veracity of the textual or the visual content itself. For this reason, fake news is also referred to as \textit{misleading content}\cite{ijmir} or \textit{fauxtography}\cite{fauxtography} in the context of social multimedia.

\subsection{Organization}
The remainder of this chapter is organized as follows. In Chapter 2, we introduce available visual features for fake news detection. We continue to present existing approaches utilizing visual content to detect fake news in Chapter 3. In Chapter 4, we discuss several challenging problems for multimedia fake news detection. Finally, we summarize available data repositories, tools (or software systems) and relevant competitions about multimedia fake news detection research in the appendix.

\section{What Visual Content Tells?}
Visual content has been shown as an important promoter for fake news propaganda\footnote{https://www.wired.com/2016/12/photos-fuel-spread-fake-news/}. At the same time, visual content also \textit{tells} abundant cues for detecting fake news. To capture the distinctive characteristics of fake news, works extracted visual features from visual content (generally, images and videos), which can be categorized into four types: forensics features, semantic features, statistical features and context features.

\subsection{Forensics Features}
Since the addressed problem is the verification of multimedia posts, one reasonable approach would be to directly verify the truth of visual content, i.e., whether the image or video is captured in the event. Intuitively, if the visual content has undergone manipulation or severe re-compression, or is generated by deep learning techniques, the news post that it belongs to is likely to be fake. To access the authenticity, (blind) forensics features which can highlight the digitally edited traces of the visual content, are exploited in fake news detection from different perspectives, including the manipulation detection, generation detection and re-compression detection.\\

\noindent \textbf{Manipulation Detection}\\
Manipulation detection aims at looking for patterns or discontinuities left by operations such as splicing, copy-move and removal. The splicing refers to copying a part of one image and inserting it into another, while the copy-move and removal both happen in the same image. Because very few works \cite{mediaeval15} directly used these features in fake news detection yet, we also investigated the features mentioned in related works and summarized as follows:

\begin{itemize}
	\item \textbf{Camera-related features} are particular patterns caused by the imaging pipeline, such as the sensor pattern noise and color filter array interpolation patterns, which can be destroyed by manipulation. In previous works, Photo-Response Non-Uniformity \cite{goljan2010defending}, noise inconsistencies \cite{mahdian2009using} and local interpolation artifacts \cite{ferrara2012image} were used to capture the change of those patterns.
	\item \textbf{Discontinuities in spatial features} are often left by forgery operations. To highlight these cues, gray-level run length features \cite{zhao2010detecting} and local binary patterns over the steerable-pyramid-transformed image \cite{muhammad2014image} were exploited.
\end{itemize}

Note that some of them are only applicable to specific types of manipulation, which is unknown in practice. Also, some widely-spread manipulated images may have undergone multiple types of processing, increasing the challenge of capturing the traces of manipulation.\\

\noindent \textbf{Generation Detection}\\
As the rapid improvement of deep generative networks (especially generative adversarial network, GAN \cite{goodfellow2014generative}), people can easily generate more photorealistic images and videos, making it hard to distinguish from natural ones. These misleading generated image and videos are often obtained by modifying the semantically-focused elements, for instance, the faces (mostly of celebrities), raising new threat to the trustworthiness of the visual content.

For generated fake images, existing works mostly focus on detecting with signal-level features. In the pixel domain, the co-occurrence matrices on three color channels were used for capturing spatial correlation characteristics, which were fed into the following convolutional neural network (CNN) for detection \cite{nataraj2019detecting}. In contrast, McCloskey et al. started with the observation in the frequency domain that GAN images have more overlapping spectral responses among the RGB channels and negative weights than natural ones \cite{mccloskey2018detecting}. To represent these differences, this work introduced intensity noise histograms and over-/under- exposed rate.

For generated fake videos, most works are devoted to the detection of DeepFakes, a series of popular implementations for superimposing existing faces onto source videos. Works for DeepFakes detection mostly focused on the local features caused by the transformation in face-swapping such as the lacking of realistic eye blinking \cite{li2018ictu}, the errors of 3D head poses introduced in face splicing for detection \cite{yang2019exposing}, and the artifacts left in warping to match the original faces \cite{li2019exposing}.\\

\noindent \textbf{Re-compression Detection} \\
A fake image or video mostly suffers multiple compression in two situations: one is that the visual content is manipulated and re-saved at last, while the other is that it is repeatedly downloaded from and uploaded to the social media platform. These two situations probably indicate deliberate manipulation of visual content or misuse of the outdated, so we can detect fake news by predicting whether the attached visual content has been re-compressed.

For images, MediaEval VMU Task \cite{mediaeval15} (see in Appendix) extracted features directly related to the compression according to \cite{dcttifs12,li2009passive}, including probability map of the aligned/non-aligned double JPEG compression, potential primary quantization steps for the first 6 Discrete Cosine Transform (DCT) coefficients of the aligned/non-aligned double JPEG compression and block artifact grid. By thresholding the aligned/non-aligned JPEG compression maps above, Boididou et al. created two binary maps considered as object and background respectively and extracted descriptive statistics (maximum, minimum, mean, median, most frequent value, standard deviation and variance) for classification \cite{boididou2015certh}. Qi et al. calculated block DCT coefficients and then performed Fourier Transform on them for enhancement to highlight the periodicity in the frequency domain caused by re-compression \cite{icdm2019}. Furthermore, because multiple spreads may cause a dramatic decrease of clarity, no-reference quality measurement \cite{wang2002no} can also indicate re-compression.

For videos, the methods exploited the presence of spikes in the Fourier transform of the energy of the displaced frame difference over time \cite{wang2006exposing}, blocking artifacts \cite{luo2008mpeg} and DCT coefficients of a macroblock \cite{wang2009exposing} to detect the double-compression (mostly in MPEG videos).

\subsection{Semantic Features}
\begin{figure}
	\vspace{-0.5cm}
	\setlength{\belowcaptionskip}{-0.5cm}
	\centering
	\subfloat[]{
		\label{pixelcase1}
		\includegraphics[height=2.1in]{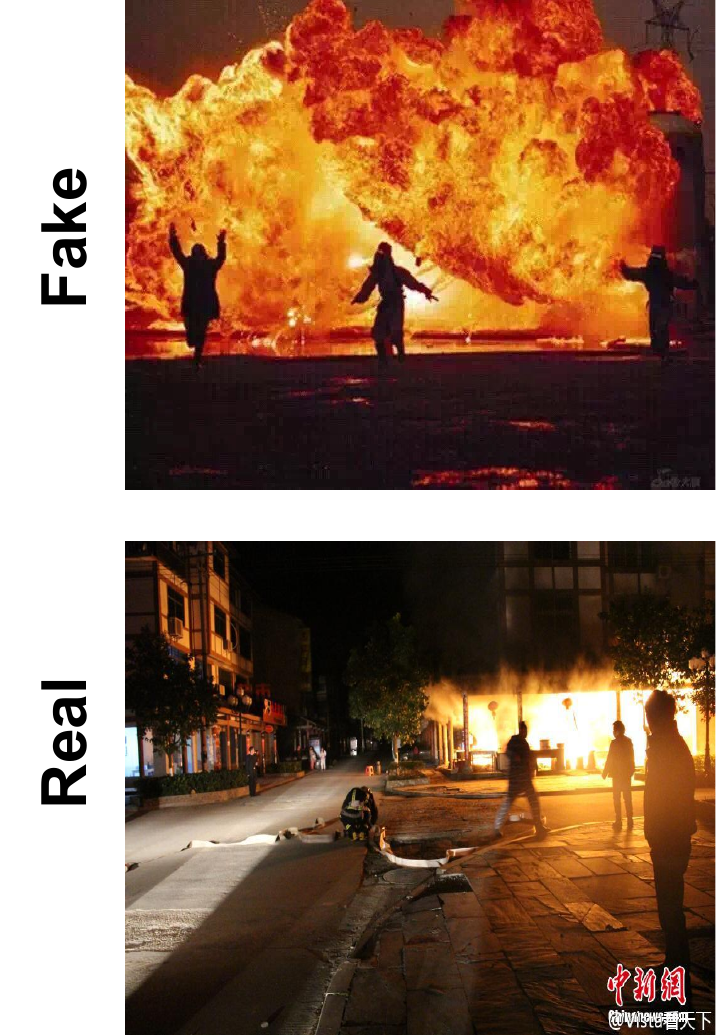}
	}
	\subfloat[]{
		\label{pixelcase1}
		\includegraphics[height=2.1in]{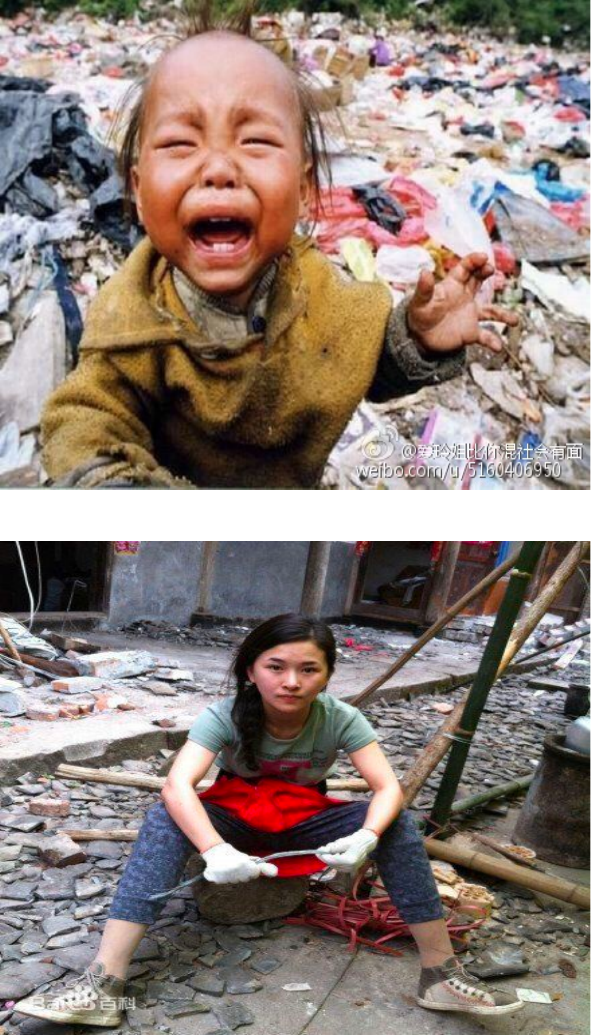}
	}
	\subfloat[]{
		\label{pixelcase3}
		\includegraphics[height=2.1in]{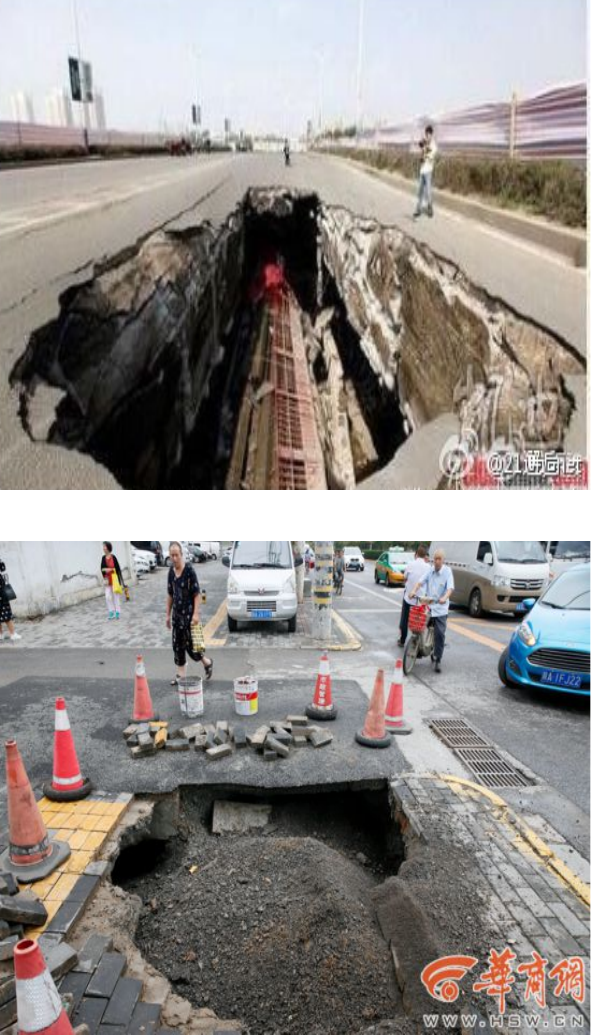}
	}\\
	\caption{Comparison of images in fake and real news images at the semantic level. We can find that fake news images are more visually striking and emotional provocative than real news images, even though they describe the same type of events such as fire (a), earthquake (b) and road collapse (c).}
	\label{Fig:semanticcase}
\end{figure}

Fake news exploits the individual vulnerabilities of people and thus often relies on sensational or even fake images to provoke anger or other emotional response of consumers for promoting the spread of fake news. Thus, images in fake news often show some distinct characteristics in comparison with real news at the semantic level, such as visual impacts \cite{jinarxiv2016image} and emotional provocations \cite{bookonrumours, surveyKai2017Fake} as Figure \ref{Fig:semanticcase} shows. Next, we introduce how to effectively extract semantic features of the visual content for fake news detection.

\begin{figure}
	\vspace{-0.5cm}
	\setlength{\belowcaptionskip}{-0.5cm}
	\centering
	\includegraphics[width=\textwidth]{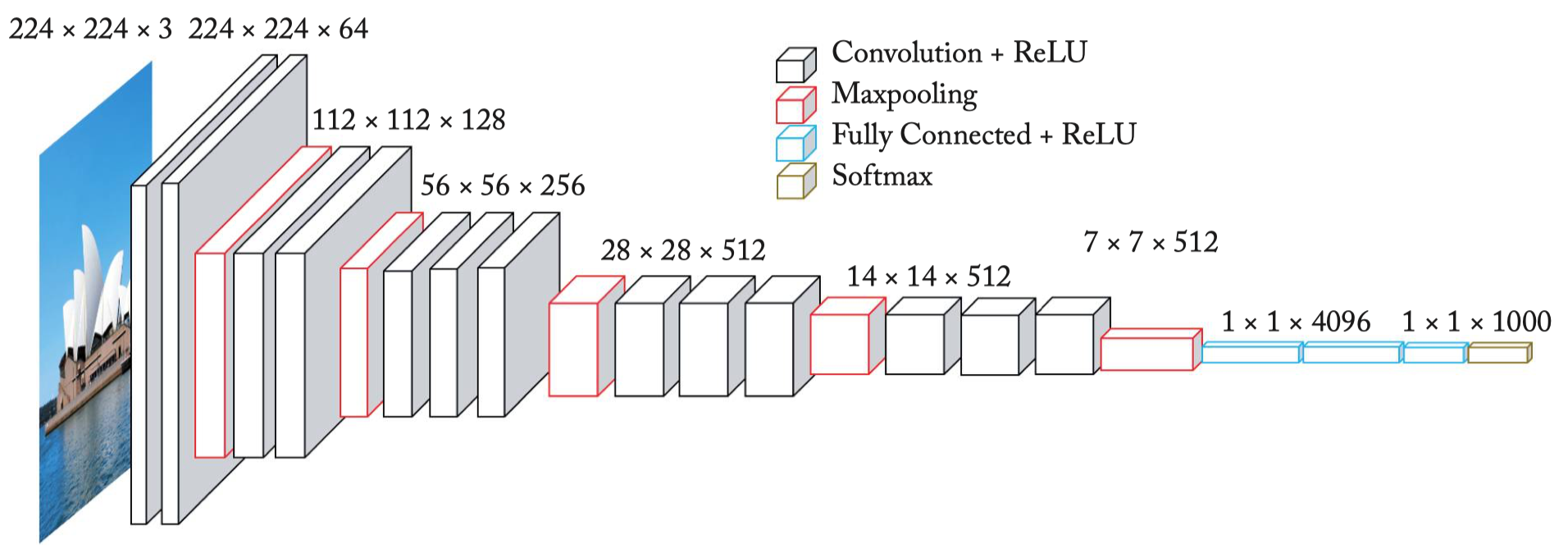}
	\caption{Detailed architecture of the VGG16 framework.}
	\label{Fig:vgg16}
\end{figure}

CNN has exhibited great power in understanding image semantics and obtaining corresponding feature representations, which can be used for various visual tasks. VGG \cite{vgg} is one of the most popular CNN models, which is comprised of three basic types of layers: convolutional layers for extracting and transforming image features, pooling layers for reducing the parameters, and fully connected layers for classification tasks (see Figure \ref{Fig:vgg16}). Most of existing works based on multimedia content adopted the VGG model to extract visual semantic features for fake news detection \cite{attRNN, eann, mvae}.

\begin{figure}
	\vspace{-0.5cm}
	\setlength{\belowcaptionskip}{-0.5cm}
	\centering
	\subfloat{
		\includegraphics[width=\textwidth]{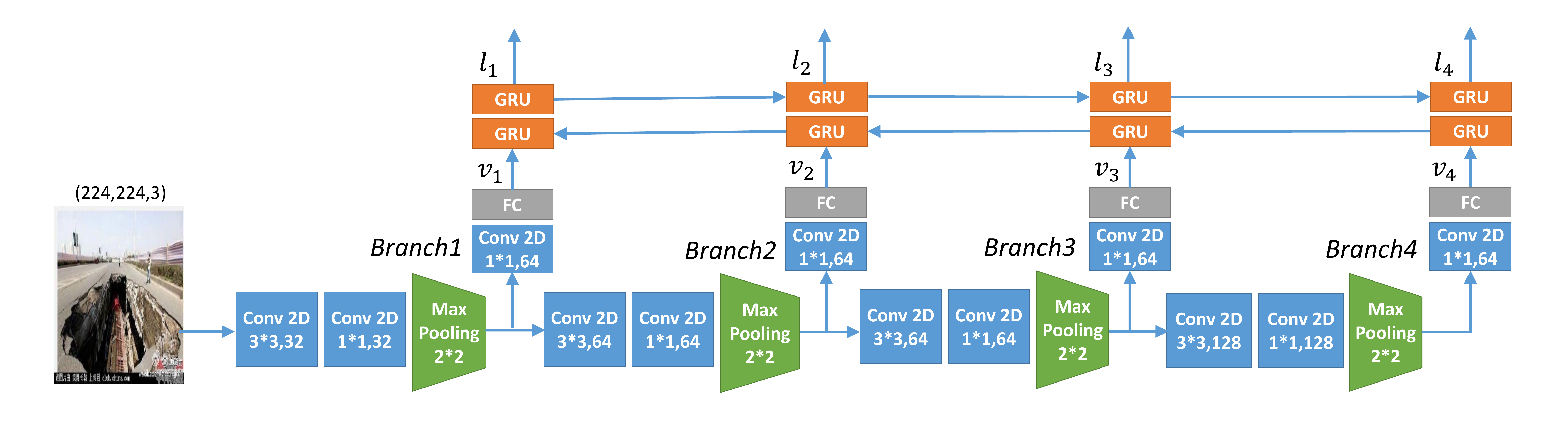}
	}\\
	\caption{Detailed architecture of the pixel domain sub-network in MVNN. For an input image, a multi-branch CNN-RNN network is utilized to extract and fuse its pixel-domain features of different semantic levels.}
	\label{Fig:pixelsubnetwork}
\end{figure}

In addition to the basic CNN, some recent works proposed novel CNN-based models to better capture the visual semantic characteristics of fake news. For example, Qi et al. proposed a multi-domain visual neural network (MVNN) to fuse the visual information of frequency and pixel domains for detecting fake news, of which the pixel sub-network was used to extract visual semantic features (see Figure \ref{Fig:pixelsubnetwork}) \cite{icdm2019}. Specifically, two motivations were illustrated for the model design. First, CNN learns high-level semantic representations through layer-by-layer abstraction from local to the global view, while the low-level features will inevitably suffer some losses in the process of abstraction. Considering these semantic cues such as emotional provocations are related to many visual factors from low-level to high-level \cite{emotionstimuli}, a multi-branch CNN network was adopted to extract features of different semantic levels in the pixel sub-network. Second, there are strong bidirectional dependencies between different levels of features. For example, middle-level features such as textures, consist of low-level features such as lines, and meanwhile compose high-level features such as objects. Therefore, the sub-network also utilized the bidirectional GRU to model the relations from two different views.

\subsection{Statistical Features}

\begin{figure}
	\vspace{-0.5cm}
	\setlength{\belowcaptionskip}{-0.5cm}
	\centering
	\subfloat[]{
		\label{statisticalcase1}
		\includegraphics[height=1.7in]{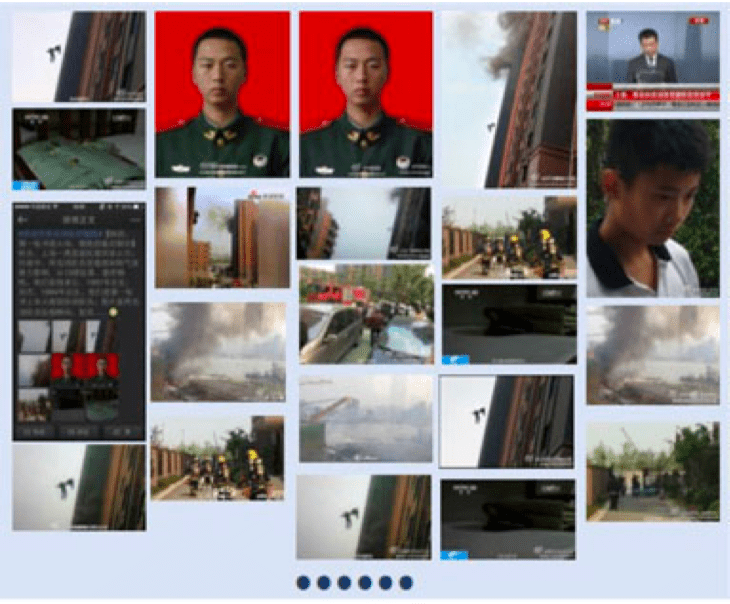}
	}
	\subfloat[]{
		\label{statisticalcase2}
		\includegraphics[height=1.7in]{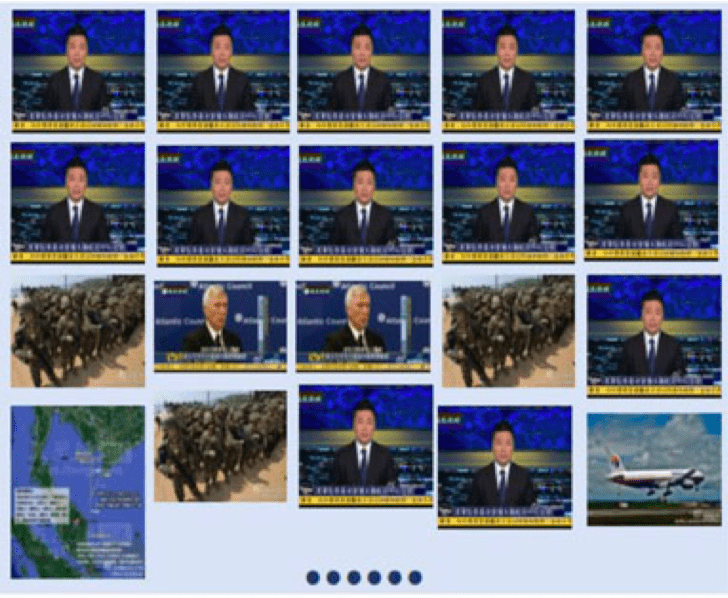}
	}\\
	\caption{Examples of images in the real and fake news event. Obviously, images in the real news event (a) are much more diverse than those in the fake one (b).}
	\label{Fig:statisticalfea}
\end{figure}

Visual content also has different distribution patterns between fake and real news on social media \cite{jin2017TMM}. Intuitively, people tend to report the news with images taken by themselves at the event scene. If the event is real, then various images taken by different witnesses would be posted while if fake, there are many repeatedly posted images with almost the same content, just as Figure \ref{Fig:statisticalfea} shows. Thus, we introduce visual statistical features to reflect this distributional difference between real and fake news.

Some works \cite{jin2017TMM, socialwu2015false, statisyang2012automatic} used basic statistical features about the attached images to assist in fake news detection, usually from three aspects:
\begin{itemize}
	\item \textbf{Count}: The occurrence number of images. For example, Wu et al. used the number of illustrations to assist detect fake news posts \cite{socialwu2015false, statisyang2012automatic}, while Jin et al. used the ratio of news posts containing at least one or more than one images to the total posts in a news event to detect fake news events \cite{jin2017TMM}.
	\item \textbf{Popularity}: The number of sharing on social media, such as re-tweets and comments. Jin et al. defined the image with a high popularity as a hot image, and regarded the ratio of hot images to all distinct images in a news event as a statistical feature \cite{jin2017TMM}.	
	\item \textbf{Type}: Some images have a particular type in resolution or style. For example, long images are images with a very large length-to-width ratio. The ratio of these types of images was also counted as a statistical feature \cite{jin2017TMM}.
\end{itemize}

In addition to these basic statistical features, Jin et al. also proposed five advanced statistical features as follows \cite{jin2017TMM}:

\begin{itemize}
	\item \textbf{Visual Clarity Score (VCS)}: Visual clarity score measures the distribution difference between two image sets: one is the image set in a certain news event (event set) and the other is the image set containing images from all events (collection set). This feature was defined as the Kullback-Leibler divergence between the two language model representing the event set and collection set, respectively. The bag-of-words image representation such as SIFT was used to define the language models for images. Specifically, the visual clarity score is
		\begin{equation}
			VCS=D_{KL}(p(w | c) \| p(w | k),
		\end{equation}
	where $p(w | c)$ and $p(w | k)$ denote the term frequency of visual word w in collection set and event set, respectively.	
	\item \textbf{Visual Coherence Score (VCoS)}: Visual coherence score measures how coherent the images in a certain news event are. This feature is computed based on the visual similarity between any image pair within images in the target event image set, which is denoted as
		\begin{equation}
			V C o S=\frac{1}{|N(N-1)|} \sum_{i, j=1, \cdots, N ; i \neq j} \textit{sim}\left(x_{i}, x_{j}\right)
		\end{equation}
	where N is number of the images in the event set, $ \textit{sim}\left(x_{i}, x_{j}\right)$ is the visual similarity between image $x_{i}$ and image $x_{j}$. In implementation, the similarity between images is computed based on their GIST features. 
	\item \textbf{Visual Similarity Distribution Histogram (VSDH)}: Visual similarity distribution histogram describes the image similarity distribution in a fine-granularity level, which is computed based on the whole similarity matrix of all images in a target news event. The similarity matrix $S$ is quantified into an $H$-bin histogram by mapping each element in the matrix into its corresponding bin, which results in a feature vector of $H$ dimensions representing the similarity relations among images,
		\begin{equation}
			VSDH(h)=\frac{1}{N^{2}}\left|\left\{(i, j) | i, j \leq N, m_{i, j} \in h-t h \operatorname{bin}\right\}\right|, h=1, \ldots, H
		\end{equation}
	\item \textbf{Visual Diversity Score (VDS)}: Visual diversity score measures the visual difference in the image set of a target news event. Assuming a ranking of images $x_{1}, x_{2}, \dots, x_{N}$ in the event image set $R$, the diversity score of all images in $R$ is,
		\begin{equation}
			\mathrm{VDS}=\sum_{i=1}^{N} \frac{1}{i} \sum_{j=1}^{i} (1-\textit{sim}\left(x_{i}, x_{j}\right))
		\end{equation}
	In implementation, images are ranked according to their popularity on social media, based on the assumption that popular images may have better representation for the news event.
	\item \textbf{Visual Clustering Score (VCS)}: Visual clustering score evaluates the image distribution over all images in the news event from a clustering perspective. It was defined as the number of clusters formed by all images in a target news event. Hierarchical agglomerative clustering (HAC) algorithm is employed to cluster these images.
	\end{itemize}

\subsection{Context Features}
According to our previous analysis, rumormongers usually use visual content from an irrelevant event to fabricate fake news. To make the fake news more reasonable, the selected visual content needs to be semantically coherent with the claim. Therefore, existing works about text-image semantic similarity aren't applicable for these manipulations. Instead, one of the most effective methods is to utilize the context information of visual content to fact-check whether the current event is the same as the original event it belongs to. Specifically, we introduce the following context features, which mainly extracted from two sources: the metadata of visual content and the external knowledge such as relevant web pages.\\

\noindent \textbf{Metadata}\\
\noindent Metadata is text information pertaining to an image/video file that is usually embedded into the file. Metadata includes not only the details relevant to the image/video itself such as file size but also the information about its production, such as position and time, which are often used in manually fact-checking \cite{2016verification,zampoglou2016web}. However, these features are not that helpful in practice because they usually become unavailable after default processing by social media.\\

\noindent \textbf{External Knowledge}\\
\noindent In addition to metadata, some works extracted context features from the external knowledge obtained through reverse image search. In contrast to classical image search, reverse image search takes an image as input and returns lo relevant web pages that include the corresponding image, title, description and time. This process can be easily automated and applied to a large number of images via some search engine APIs like google reverse image search\footnote{https://images.google.com/}. Next, we introduce three context features as follows.

\begin{itemize}
	\item \textbf{Timespan}: Timespan is defined as the time delay between the published time of the news and the earliest published time of the visual content. This feature is proposed to verify the originality of the visual content \cite{timespan}. If the timespan is bigger than a specific threshold, then the visual content is probably from an irrelevant event.
	\item \textbf{Inter-claim similarity}: Inter-claim similarity is defined as the similarity between the claim and the textual contents of these crawled websites. Considering that the text information of these crawled websites is helpful for understanding the original event of the image, this feature is used to verify the event consistency between the textual claim and corresponding visual content \cite{emnlp2019}.
	\item \textbf{Platform credibility}: Platform credibility means the credibility of the source platform where the visual content was published \cite{emnlp2019}. By using the dataset of Media Bias/Fact Check (MBFC)\footnote{http://mediabiasfactcheck.com/}, a web site that provides factuality information about 2700+ media sources, each web page that is returned by the reverse image search was classified into the following categories: high factuality, low factuality and mixed factuality. The percentage of web pages from each category returned by the reverse image search was defined as the platform credibility feature.
\end{itemize}

\section{How Visual Content Helps?}

In the previous section, we introduced four types of visual features from different perspectives, i.e., forensics features, semantic features, statistical features and context features, for multimedia fake news detection. These features reflect the characteristics of visual content and are usually combined in practice for covering more situations. In this section, we discuss the details of several existing approaches utilizing visual content to detect fake news, which can be broadly classified into content-based approaches and knowledge-based approaches. \textbf{Content-based approaches} focus on capturing and combining the cues from contents of different modalities for fake news detection, without using any reference datasets. \textbf{Knowledge-based approaches} aim to use external sources to fact-check input claims. They assume the existence of a relatively large reference dataset and assess the integrity of the news post by comparing it to one or more posts retrieved from the reference dataset.

\subsection{Content-based Approaches}
A complete news story consists of textual and visual content simultaneously, which both provide distinctive cues for detecting fake news. Therefore, recent works on this problem focus on utilize and effectively fuse information from multiple modalities. Mostly, these works simply used a common recurrent neural network (RNN) and a pre-trained CNN to obtain the textual and visual semantic features. Next, we introduce three state-of-the-art approaches that fuse multimodal information for fake news detection.

\begin{figure}
	\vspace{-0.5cm}
	\setlength{\belowcaptionskip}{-0.5cm}
	\centering
	\subfloat[]{
		\label{attRNN}
		\includegraphics[width=0.5\textwidth,height=1.4in]{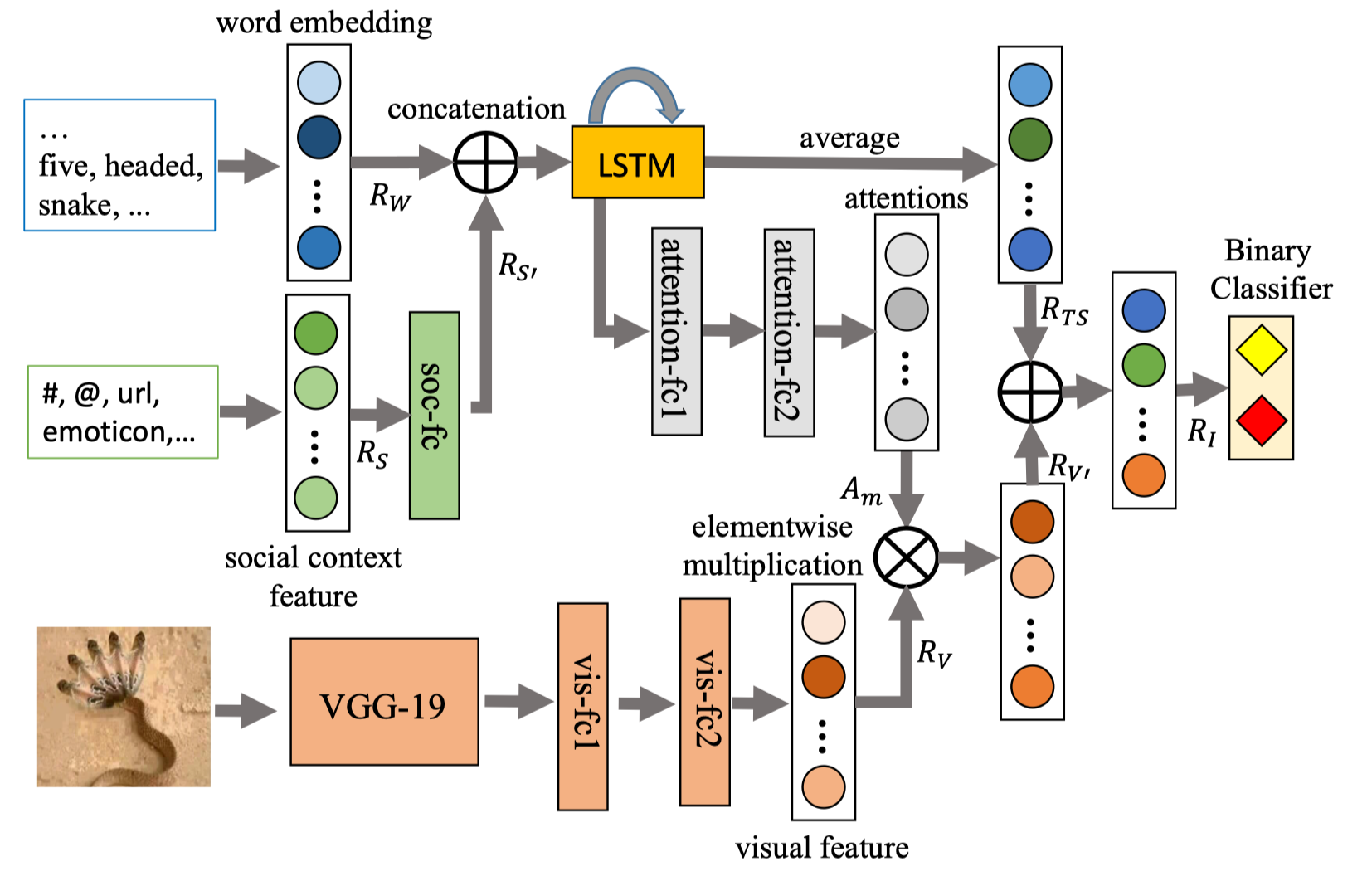}
	}
	\subfloat[]{
		\label{eann}
		\includegraphics[width=0.5\textwidth,height=1.4in]{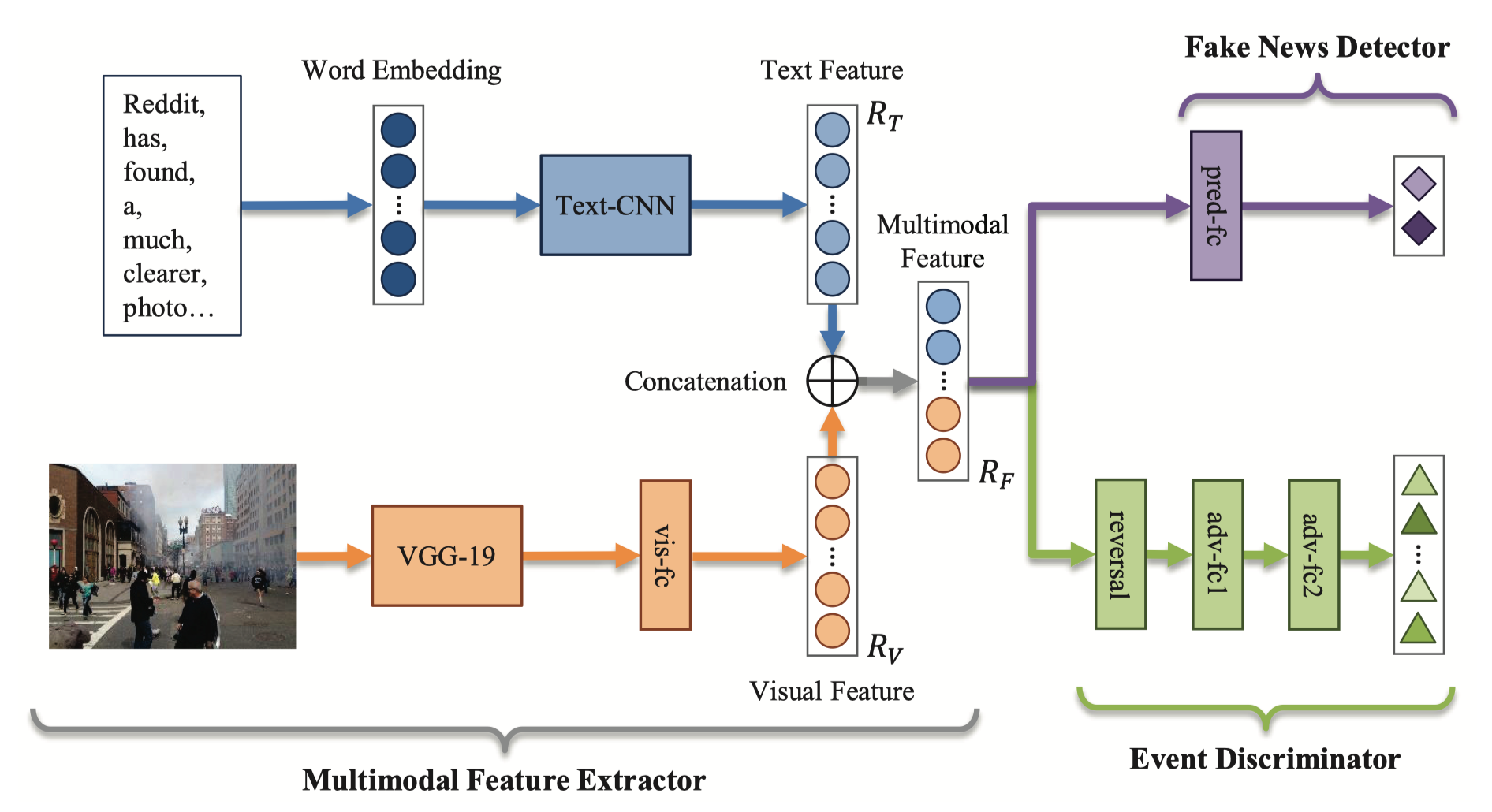}
	}\\
	\subfloat[]{
		\label{mvae}
		\includegraphics[height=1.6in]{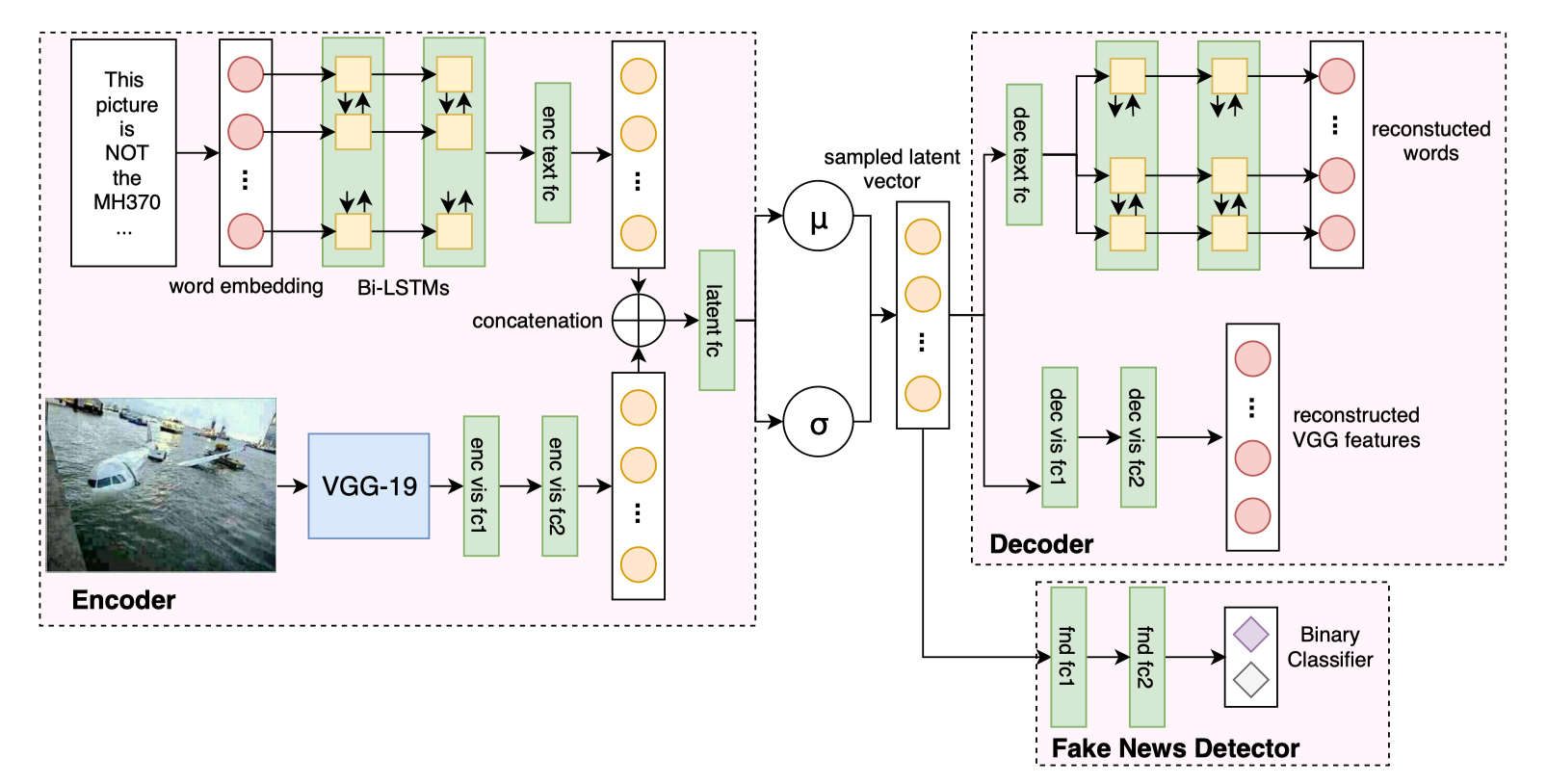}
	}\\
	\caption{Architectures of three state-of-the-art multi-modal models for fake news detection. (a) attRNN. (b) EANN. (c) MVAE.}
	\label{Fig:models}
\end{figure}

Jin et al. \cite{attRNN} first incorporated multi-modal contents via deep neural networks to solve fake news detection problem. It proposed an innovative RNN with an attention mechanism (attRNN, see Figure \ref{attRNN}) for effectively fusing the textual, visual and social context features. For a given tweet, its text and social context are first fused with an LSTM for a joint representation. This representation is then fused with visual features extracted from pre-trained deep CNN. The output of the LSTM at each time step is employed as the neuron-level attention to coordinate visual features during the fusion.

Wang et al. \cite{eann} proposed an end-to-end event adversarial neural network (EANN, see Figure \ref{eann}) to detect newly-emerged fake news events based on event-invariant multi-modal features. It consists of three main components: the multi-modal feature extractor, the fake news detector, and the event discriminator. The multi-modal feature extractor is responsible for extracting the textual and visual features from posts. It cooperates with the fake news detector to learn the discriminable representation for fake news detection. The role of event discriminator is to remove the event-specific features and keep shared features among events.

Dhruv et al. \cite{mvae} utilized a multi-modal variational autoencoder (MVAE, see Figure \ref{mvae}) trained jointly with a fake news detector to learn a shared representation of textual and visual information. The model consists of three main components: an encoder, a decoder and a fake news detector module. The variational autoencoder is capable of learning probabilistic latent variable models by optimizing a bound on the marginal likelihood of the observed data. The fake news detector then utilizes the multi-modal representations obtained from the bi-modal variational autoencoder to classify posts as fake or not.

\subsection{Knowledge-based Approaches}
Real-world multimedia news is often composed of multiple modalities, like the image or a video with associated text and metadata, where information about an event is incompletely captured by each modality separately. Such multimedia data packages, i.e., the tuples of multi-modal information of the posts, are prone to manipulations, where a subset of these modalities can be modified to misrepresent or repurpose the multimedia package. However, the details being manipulated are subtle and often interleaved with the truth, causing that the content-based approaches can hardly detect these manipulations. Faced with this problem, knowledge-based approaches utilize external sources, a reference dataset of unmanipulated packages as a source of world knowledge, to help verify the semantic integrity of the multimedia news. In the following, we introduce some representative knowledge-based methods.

\begin{figure}
	\vspace{-0.5cm}
	\setlength{\belowcaptionskip}{-0.5cm}
	\centering
	\includegraphics[width=0.85\textwidth]{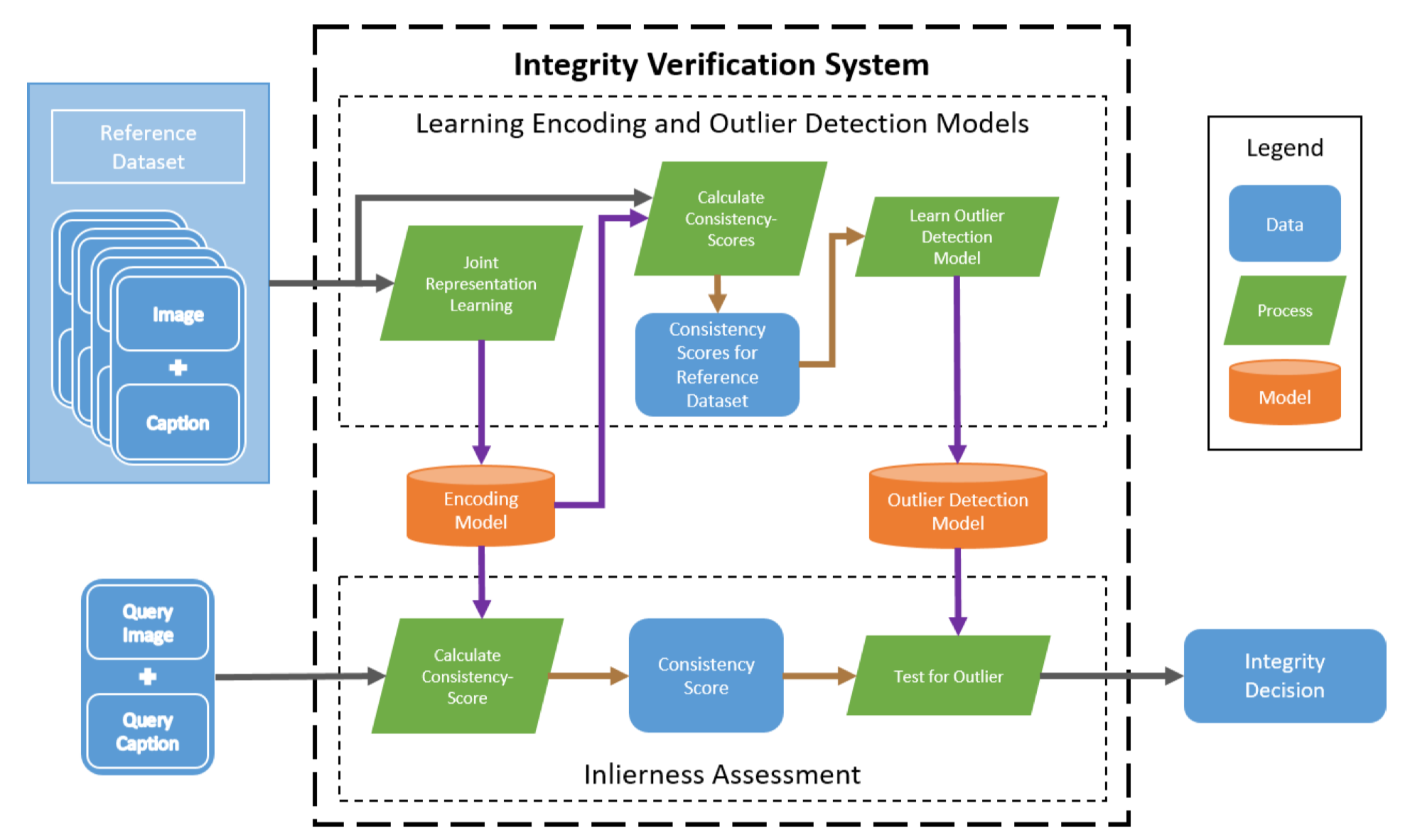}
	\caption{The package integrity assessment system of \cite{knowledgemm17}.}
	\label{Fig:mm17}
\end{figure}

Jaiswal et al. \cite{knowledgemm17} first formally defined the multimedia semantic integrity assessment problem and combined deep multi-modal representation learning with outlier detection methods to assess whether a caption was consistent with the image in its package (see Figure \ref{Fig:mm17}). Data packages in the reference dataset were used to train a deep multi-modal representation learning model, which was then used to assess the integrity of query packages by calculating image-caption consistency scores and employing outlier detection models to find their inlierness with respect to the reference dataset.

\begin{figure}
	\vspace{-0.5cm}
	\setlength{\belowcaptionskip}{-0.5cm}
	\centering
	\includegraphics[width=\textwidth]{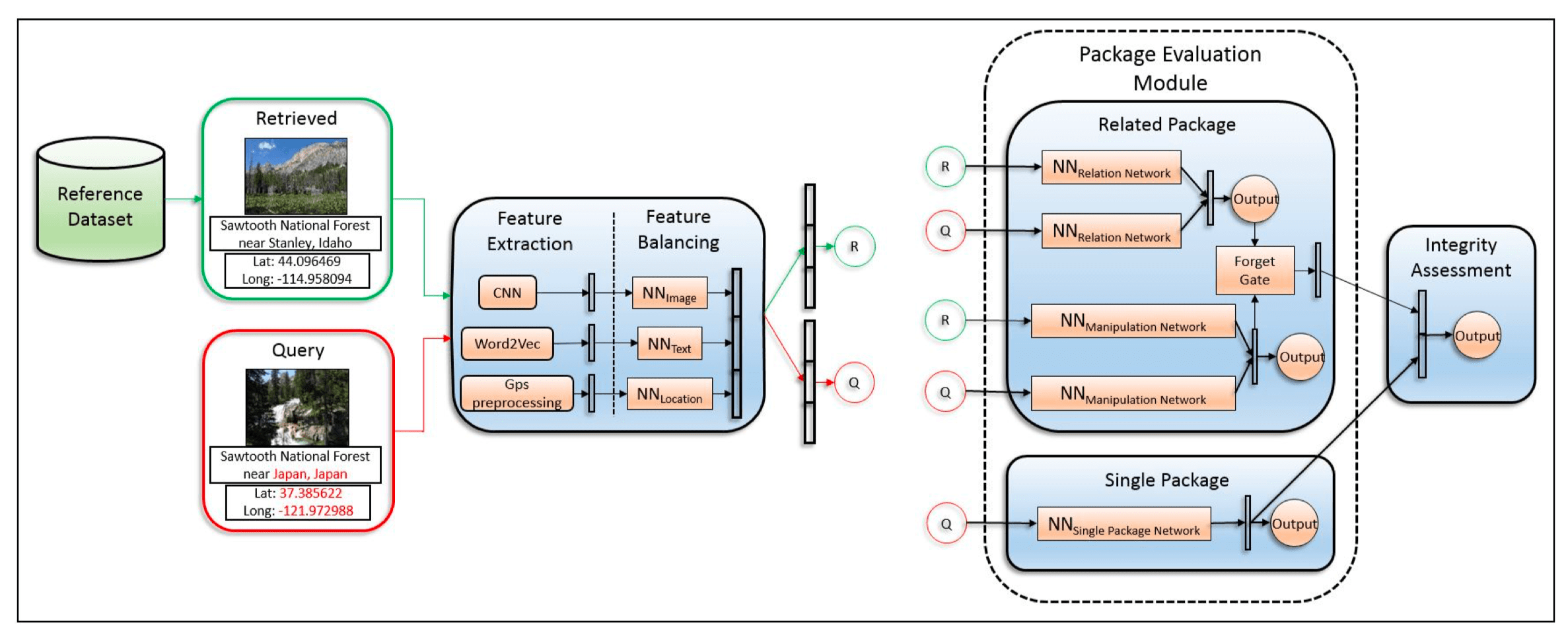}
	\caption{The Package Integrity Assessment Model of \cite{knowledgemm18}.}
	\label{Fig:mm18}
\end{figure}

Similarly, Sabir et al. \cite{knowledgemm18} proposed a novel deep multi-modal model (see Figure \ref{Fig:mm18}) to verify the integrity of multimedia packages. The proposed model consists of four modules: (1) feature extraction, (2) feature balancing, (3) package evaluation and (4) integrity assessment. For each query package, the model first uses similarity scoring to retrieve a package from the reference dataset, taking the query package and the top-1 related package as the input of the model. After passing to the feature extraction and balancing modules, query and retrieved packages are transformed into a single feature vector. The package evaluation module, the core of the proposed model, consists of the related package and single package sub-modules. The related package sub-module consisted of two siamese networks. The first network is a relationship classifier that verifies whether the query package and top-1 package are indeed related, while the second network is a manipulation detector that determines whether the query package is a manipulated version of the top-1 retrieved package. Since manipulation detection is dependent on the relatedness of the two packages, the relationship classifier controls a forget gate which scales the feature vector of the manipulation detector according to the relatedness between the two packages. In the meantime, a single package module verifies the coherency (i.e., integrity) of the query package alone. The integrity assessment module concatenated feature vectors from both related and single package modules for manipulation classification.

\begin{figure}
	\vspace{-0.5cm}
	\setlength{\belowcaptionskip}{-0.5cm}
	\centering
	\includegraphics[width=.7\textwidth]{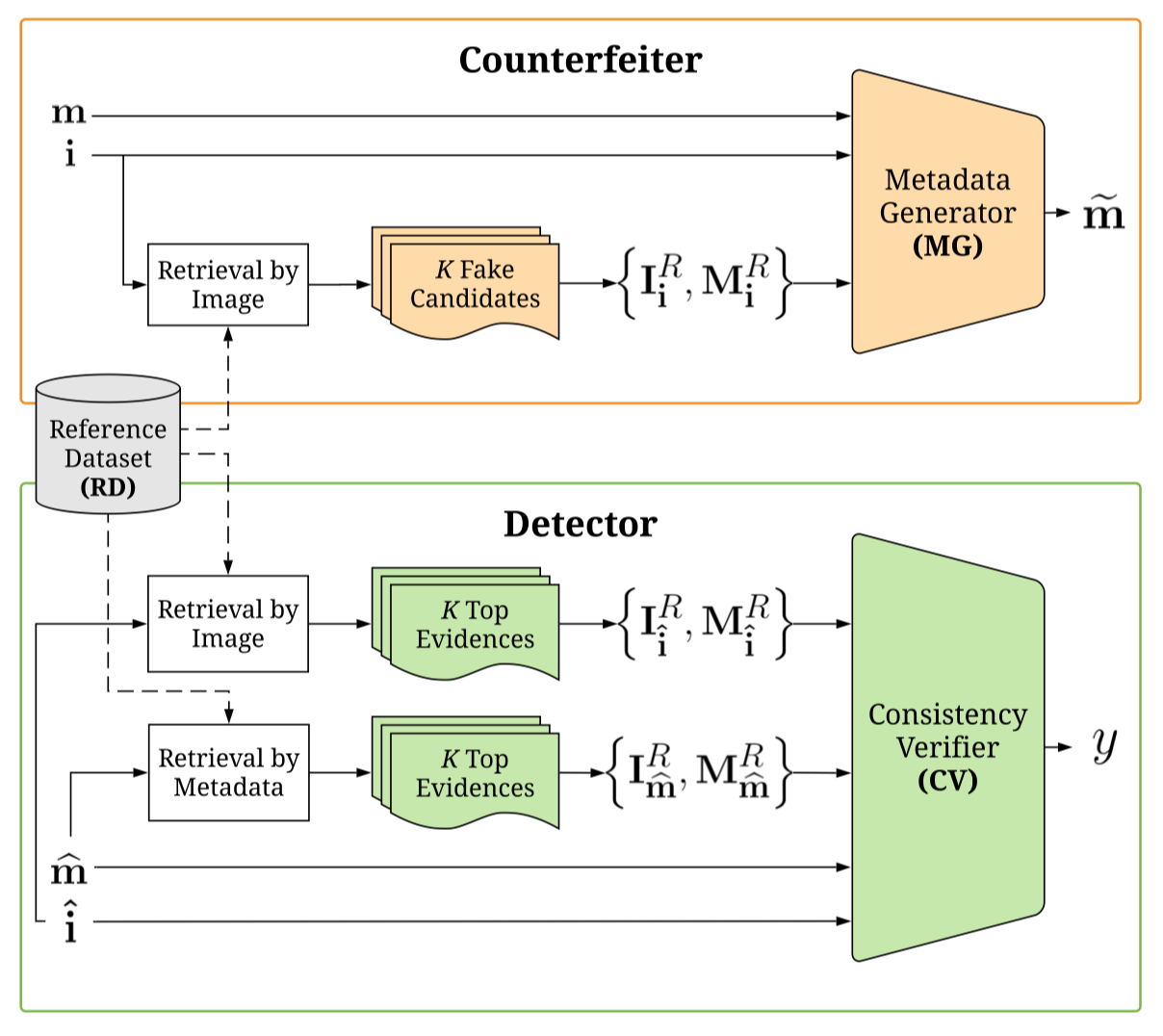}
	\caption{Architecture of Adversarial Image Repurposing Detection (AIRD).}
	\label{Fig:cvpr19}
\end{figure}

One of the main challenges for developing multimedia semantic integrity assessment methods is the lack of training and evaluation data. In light of this, Jaiswal et al. \cite{knowledgecvpr19} proposed a novel framework, Adversarial Image Repurposing Detection (AIRD) (see Figure \ref{Fig:cvpr19}), for image repurposing detection, which can be trained in the absence of training data containing manipulated metadata. AIRD is to simulate the real-world adversarial interplay between a bad actor who repurposes images with counterfeit metadata and a watchdog who verifies the semantic consistency between images and their accompanying metadata. More specifically, AIRD consists of two models: a counterfeiter and a detector, which are trained in an adversarial way. While the detector gathers evidence from the reference set, the counterfeiter exploits it to conjure convincingly deceptive fake metadata for a given query package.

\section{Challenging Problems}
In the previous sections, we introduce several visual features and existing approaches based on visual content for effective fake news detection. Despite the research developments on the multimedia fake news detection problem, there are still some specific challenges that need to be considered.

One major challenge is the lacking of labeled data. Although the multimedia content is rapidly growing nowadays, datasets about multimedia fake news are scarce, which hinders the development of this research field. To tackle this challenge, on the one hand, we encourage researchers to pay more attention to constructing and releasing high-quality labeled datasets. On the other hand, it is important to study multimedia fake news detection in a weakly supervised setting, i.e., with limited or no label data for training. For example, Jin et al. \cite{jinarxiv2016image} constructs a large-scale weakly-labeled dataset as auxiliary to overcome the data scarcity issue, and proposes a domain transferred deep CNN to detect the fake news images.

Besides, another critical challenge is the explainability of fake news detection, i.e., why a model determines a particular piece of news as fake. Although computational detection of fake news has produced some promising results, the explainability of such detection remains largely unsolved, making the judgments unconvincing. In recent years, fact-checking approaches have aroused the attention of researchers, which could offer a new way to tackle this challenge. Different from traditional style-based fake news detection, these approaches utilize external resources (also known as knowledge) as evidence to fact-check a given piece of news is fake or real. For multimedia content, the relationship between the textual and visual content and metadata is a powerful clue, which can be combined with the external knowledge to make inferences. These approaches are helpful for better understanding and explaining the decision made by algorithms according to the involved evidence and visible inference process.\\

\noindent \textbf{Acknowledgements.}
 This work was supported by the National Natural Science Foundation of China (U1703261).

%
%
%
 \bibliographystyle{splncs04}
 \bibliography{chapter}

\begin{thebibliography}{10}
\providecommand{\url}[1]{\texttt{#1}}
\providecommand{\urlprefix}{URL }
\providecommand{\doi}[1]{https://doi.org/#1}

\bibitem{fakenewsdefinition}
Allcott, H., Gentzkow, M.: Social media and fake news in the 2016 election.
  Journal of Economic Perspectives  \textbf{31}(2),  211--36 (2017)

\bibitem{dcttifs12}
Bianchi, T., Piva, A.: Image forgery localization via block-grained analysis of
  jpeg artifacts. IEEE Transactions on Information Forensics and Security
  \textbf{7}(3),  1003--1017 (2012)

\bibitem{mediaeval15}
Boididou, C., Andreadou, K., Papadopoulos, S., Dang-Nguyen, D.T., Boato, G.,
  Riegler, M., Kompatsiaris, Y., et~al.: Verifying multimedia use at mediaeval
  2015. In: MediaEval (2015)

\bibitem{boididou2015certh}
Boididou, C., Papadopoulos, S., Dang-Nguyen, D.T., Boato, G., Kompatsiaris, Y.:
  The certh-unitn participation@ verifying multimedia use 2015. In: MediaEval
  (2015)

\bibitem{mediaeval16}
Boididou, C., Papadopoulos, S., Dang-Nguyen, D.T., Boato, G., Riegler, M.,
  Middleton, S.E., Petlund, A., Kompatsiaris, Y., et~al.: Verifying multimedia
  use at mediaeval 2016. In: MediaEval (2016)

\bibitem{ijmir}
Boididou, C., Papadopoulos, S., Zampoglou, M., Apostolidis, L., Papadopoulou,
  O., Kompatsiaris, Y.: Detection and visualization of misleading content on
  twitter. International Journal of Multimedia Information Retrieval
  \textbf{7}(1),  71--86 (2018)

\bibitem{2016verification}
Brandtzaeg, P.B., L{\"u}ders, M., Spangenberg, J., Rath-Wiggins, L.,
  F{\o}lstad, A.: Emerging journalistic verification practices concerning
  social media. Journalism Practice  \textbf{10}(3),  323--342 (2016)

\bibitem{cao2019false}
Cao, J., Sheng, Q., Qi, P., Zhong, L., Wang, Y., Zhang, X.: False news
  detection on social media. arXiv preprint arXiv:1908.10818  (2019)

\bibitem{mvae}
Dhruv, K., Jaipal~Singh, G., Manish, G., Vasudeva, V.: Mvae: Multimodal
  variational autoencoder for fake news detection. In: Proceedings of the 2019
  World Wide Web Conference. ACM (2019)

\bibitem{ferrara2012image}
Ferrara, P., Bianchi, T., De~Rosa, A., Piva, A.: Image forgery localization via
  fine-grained analysis of cfa artifacts. IEEE Transactions on Information
  Forensics and Security  \textbf{7}(5),  1566--1577 (2012)

\bibitem{goljan2010defending}
Goljan, M., Fridrich, J., Chen, M.: Defending against fingerprint-copy attack
  in sensor-based camera identification. IEEE Transactions on Information
  Forensics and Security  \textbf{6}(1),  227--236 (2010)

\bibitem{goodfellow2014generative}
Goodfellow, I., Pouget-Abadie, J., Mirza, M., Xu, B., Warde-Farley, D., Ozair,
  S., Courville, A., Bengio, Y.: Generative adversarial nets. In: Advances in
  neural information processing systems. pp. 2672--2680 (2014)

\bibitem{knowledgemm17}
Jaiswal, A., Sabir, E., AbdAlmageed, W., Natarajan, P.: Multimedia semantic
  integrity assessment using joint embedding of images and text. In:
  Proceedings of the 25th ACM international conference on Multimedia. pp.
  1465--1471. ACM (2017)

\bibitem{knowledgecvpr19}
Jaiswal, A., Wu, Y., AbdAlmageed, W., Masi, I., Natarajan, P.: Aird:
  Adversarial learning framework for image repurposing detection. In:
  Proceedings of the IEEE Conference on Computer Vision and Pattern
  Recognition. pp. 11330--11339 (2019)

\bibitem{attRNN}
Jin, Z., Cao, J., Guo, H., Zhang, Y., Luo, J.: Multimodal fusion with recurrent
  neural networks for rumor detection on microblogs. In: Proceedings of the
  2017 ACM on Multimedia Conference. pp. 795--816. ACM (2017)

\bibitem{jinarxiv2016image}
Jin, Z., Cao, J., Luo, J., Zhang, Y.: Image credibility analysis with effective
  domain transferred deep networks. arXiv preprint arXiv:1611.05328  (2016)

\bibitem{jin2017TMM}
Jin, Z., Cao, J., Zhang, Y., Zhou, J., Tian, Q.: Novel visual and statistical
  image features for microblogs news verification. IEEE Transactions on
  Multimedia  \textbf{19}(3),  598--608 (2017)

\bibitem{surveykumar2018false}
Kumar, S., Shah, N.: False information on web and social media: A survey. arXiv
  preprint arXiv:1804.08559  (2018)

\bibitem{emotionstimuli}
Lang, P.J.: A bio-informational theory of emotional imagery. Psychophysiology
  \textbf{16}(6),  495--512 (1979)

\bibitem{thescienceoffakenews}
Lazer, D.M., Baum, M.A., Benkler, Y., Berinsky, A.J., Greenhill, K.M., Menczer,
  F., Metzger, M.J., Nyhan, B., Pennycook, G., Rothschild, D., et~al.: The
  science of fake news. Science  \textbf{359}(6380),  1094--1096 (2018)

\bibitem{li2009passive}
Li, W., Yuan, Y., Yu, N.: Passive detection of doctored jpeg image via block
  artifact grid extraction. Signal Processing  \textbf{89}(9),  1821--1829
  (2009)

\bibitem{li2018ictu}
Li, Y., Chang, M.C., Lyu, S.: In ictu oculi: Exposing ai generated fake face
  videos by detecting eye blinking. arXiv preprint arXiv:1806.02877  (2018)

\bibitem{li2019exposing}
Li, Y., Lyu, S.: Exposing deepfake videos by detecting face warping artifacts.
  In: Proceedings of the IEEE Conference on Computer Vision and Pattern
  Recognition Workshops. pp. 46--52 (2019)

\bibitem{luo2008mpeg}
Luo, W., Wu, M., Huang, J.: Mpeg recompression detection based on block
  artifacts. In: Security, Forensics, Steganography, and Watermarking of
  Multimedia Contents X. vol.~6819, p. 68190X. International Society for Optics
  and Photonics (2008)

\bibitem{ma2016detecting}
Ma, J., Gao, W., Mitra, P., Kwon, S., Jansen, B.J., Wong, K.F., Cha, M.:
  Detecting rumors from microblogs with recurrent neural networks. In:
  Proceedings of the Twenty-Fifth International Joint Conference on Artificial
  Intelligence. pp. 3818--3824. AAAI Press (2016)

\bibitem{mahdian2009using}
Mahdian, B., Saic, S.: Using noise inconsistencies for blind image forensics.
  Image and Vision Computing  \textbf{27}(10),  1497--1503 (2009)

\bibitem{mccloskey2018detecting}
McCloskey, S., Albright, M.: Detecting gan-generated imagery using color cues.
  arXiv preprint arXiv:1812.08247  (2018)

\bibitem{muhammad2014image}
Muhammad, G., Al-Hammadi, M.H., Hussain, M., Bebis, G.: Image forgery detection
  using steerable pyramid transform and local binary pattern. Machine Vision
  and Applications  \textbf{25}(4),  985--995 (2014)

\bibitem{nataraj2019detecting}
Nataraj, L., Mohammed, T.M., Manjunath, B., Chandrasekaran, S., Flenner, A.,
  Bappy, J.H., Roy-Chowdhury, A.K.: Detecting gan generated fake images using
  co-occurrence matrices. arXiv preprint arXiv:1903.06836  (2019)

\bibitem{icdm2019}
Qi, P., Cao, J., Yang, T., Guo, J., Li, J.: Exploiting multi-domain visual
  information for fake news detection. In: 19th IEEE International Conference
  on Data Mining. IEEE (2019)

\bibitem{knowledgemm18}
Sabir, E., AbdAlmageed, W., Wu, Y., Natarajan, P.: Deep multimodal
  image-repurposing detection. In: 2018 ACM Multimedia Conference on Multimedia
  Conference. pp. 1337--1345. ACM (2018)

\bibitem{fakenewsnet}
Shu, K., Mahudeswaran, D., Wang, S., Lee, D., Liu, H.: Fakenewsnet: A data
  repository with news content, social context and dynamic information for
  studying fake news on social media. arXiv preprint arXiv:1809.01286  (2018)

\bibitem{surveyKai2017Fake}
Shu, K., Sliva, A., Wang, S., Tang, J., Liu, H.: Fake news detection on social
  media: A data mining perspective. ACM SIGKDD Explorations Newsletter
  \textbf{19}(1),  22--36 (2017)

\bibitem{vgg}
Simonyan, K., Zisserman, A.: Very deep convolutional networks for large-scale
  image recognition. arXiv preprint arXiv:1409.1556  (2014)

\bibitem{timespan}
Sun, S., Liu, H., He, J., Du, X.: Detecting event rumors on sina weibo
  automatically. In: Asia-Pacific Web Conference. pp. 120--131. Springer (2013)

\bibitem{bookonrumours}
Sunstein, C.R.: On rumors: How falsehoods spread, why we believe them, and what
  can be done. Princeton University Press (2014)

\bibitem{wang2006exposing}
Wang, W., Farid, H.: Exposing digital forgeries in video by detecting double
  mpeg compression. In: Proceedings of the 8th workshop on Multimedia and
  security. pp. 37--47. ACM (2006)

\bibitem{wang2009exposing}
Wang, W., Farid, H.: Exposing digital forgeries in video by detecting double
  quantization. In: Proceedings of the 11th ACM workshop on Multimedia and
  security. pp. 39--48. ACM (2009)

\bibitem{wang2017liar}
Wang, W.Y.: “liar, liar pants on fire”: A new benchmark dataset for fake
  news detection. In: Proceedings of the 55th Annual Meeting of the Association
  for Computational Linguistics (Volume 2: Short Papers). pp. 422--426 (2017)

\bibitem{eann}
Wang, Y., Ma, F., Jin, Z., Yuan, Y., Xun, G., Jha, K., Su, L., Gao, J.: Eann:
  Event adversarial neural networks for multi-modal fake news detection. In:
  Proceedings of the 24th ACM SIGKDD International Conference on Knowledge
  Discovery \& Data Mining. pp. 849--857. ACM (2018)

\bibitem{wang2002no}
Wang, Z., Sheikh, H.R., Bovik, A.C.: No-reference perceptual quality assessment
  of jpeg compressed images. In: Proceedings. International Conference on Image
  Processing. vol.~1, pp.~I--I. IEEE (2002)

\bibitem{socialwu2015false}
Wu, K., Yang, S., Zhu, K.Q.: False rumors detection on sina weibo by
  propagation structures. In: 2015 IEEE 31st International Conference on Data
  Engineering. pp. 651--662. IEEE (2015)

\bibitem{statisyang2012automatic}
Yang, F., Liu, Y., Yu, X., Yang, M.: Automatic detection of rumor on sina
  weibo. In: Proceedings of the ACM SIGKDD Workshop on Mining Data Semantics.
  p.~13. ACM (2012)

\bibitem{yang2019exposing}
Yang, X., Li, Y., Lyu, S.: Exposing deep fakes using inconsistent head poses.
  In: ICASSP 2019-2019 IEEE International Conference on Acoustics, Speech and
  Signal Processing (ICASSP). pp. 8261--8265. IEEE (2019)

\bibitem{zampoglou2016web}
Zampoglou, M., Papadopoulos, S., Kompatsiaris, Y., Bouwmeester, R.,
  Spangenberg, J.: Web and social media image forensics for news professionals.
  In: Tenth International AAAI Conference on Web and Social Media (2016)

\bibitem{fauxtography}
Zhang, D.Y., Shang, L., Geng, B., Lai, S., Li, K., Zhu, H., Amin, M.T., Wang,
  D.: Fauxbuster: A content-free fauxtography detector using social media
  comments. In: 2018 IEEE International Conference on Big Data (Big Data). pp.
  891--900. IEEE (2018)

\bibitem{zhao2010detecting}
Zhao, X., Li, J., Li, S., Wang, S.: Detecting digital image splicing in chroma
  spaces. In: International Workshop on Digital Watermarking. pp. 12--22.
  Springer (2010)

\bibitem{emnlp2019}
Zlatkova, D., Nakov, P., Koychev, I.: Fact-checking meets fauxtography:
  Verifying claims about images. In: Proceedings of the 2019 Conference on
  Empirical Methods in Natural Language Processing and the 9th International
  Joint Conference on Natural Language Processing (EMNLP-IJCNLP). pp.
  2099--2108 (2019)

\bibitem{surveyzubiaga2018}
Zubiaga, A., Aker, A., Bontcheva, K., Liakata, M., Procter, R.: Detection and
  resolution of rumours in social media: A survey. ACM Computing Surveys (CSUR)
   \textbf{51}(2), ~32 (2018)

\end{thebibliography}
%
%
%
%
%

\section*{Appendix}
\subsection*{Data Repositories}
The step above all to detect fake news is to collect a real-world benchmark dataset. Though several text-based fake news datasets \cite{ma2016detecting, wang2017liar} have been released, publicized multimedia fake news datasets remain rare, hindering the development of fake multimedia news detection. We here introduce representative multimedia datasets in fake news detection as follows.

\textbf{MediaEval-VMU}\footnote{https://github.com/MKLab-ITI/image-verification-corpus}: The earliest publicly available multimedia verification corpus originates from the MediaEval 2015 Verifying Multimedia Use (VMU) task \cite{mediaeval15}, which is further extended in 2016 \cite{mediaeval16}. In the latest version, the dataset consists of tweets from Twitter related to 17 events (or hoaxes) that comprise in total 193 cases of real images, 218 cases of misused (fake) images and two cases of misused videos, associated with 6,225 real and 9,404 fake tweets posted by 5,895 and 9,025 unique users, respectively.

\textbf{TMM17}: Due to the insufficiency of images in previous works like VMU, Jin et al. \cite{jin2017TMM} collected a new dataset by crawling posts related to the authoritatively verified events from Weibo. The dataset is constituted of 146 news events with 50,287 posts posted by 42,441 distinct users. A total of 25,953 images are attached to 19,762 of the posts. Note that this work focuses on event-level detection, so there exist posts with no image attached.

\textbf{MM17}\cite{attRNN}: This multimedia dataset is especially for multi-modal fake news detection. The authors used similar sources as \cite{jin2017TMM}, but text-only posts and posts with duplicated, small-size and large-height images were removed. The dataset finally consists of 9,528 posts, with balanced amounts of fake and real news.

\textbf{FakeNewsNet}\footnote{https://github.com/KaiDMML/FakeNewsNet}: In \cite{fakenewsnet}, Shu et al. collected fake news articles instead of short statements by traversing the fact-check websites such as PolitiFact\footnote{https://www.politifact.com/} and GossipCop\footnote{https://www.gossipcop.com/} and then searching for the web pages of corresponding articles. Totally, 336 fake and 447 real news articles contain images in PolitiFact part, while 1,650 fake and 16,767 real do in GossipCop part.

\textbf{MCG-FNeWS}\footnote{http://mcg.ict.ac.cn/wordpress/share/mcg-fnews/}\cite{cao2019false}: The first version of this dataset was released for the False News Detection Competition 2019. The data was collected from Weibo official debunking center\footnote{https://service.account.weibo.com/} and news verification system AI-Shiyao\footnote{ https://www.newsverify.com/} and reorganized for different sub-tasks in the competition. For multi-modal detection sub-task, the whole set consists of 46,373 posts (23,186 real and 23,187 fake) with 41,937 images (24,794 in real posts and 17,170 in fake posts).

\textbf{EMNLP19}\footnote{http://gitlab.com/didizlatkova/fake-image-detection} \cite{emnlp2019}: This dataset is especially for verifying the claims about images. The image-related news was collected from two sources: A section of Snopes.com named \textit{Fauxtography}\footnote{https://www.snopes.com/fact-check/category/photos/} for all false image-related news and a small fraction of true news; Reuters' \textit{Picture of the Year} from 2015 to 2018 for most of true news. In total, this dataset contains 592 true and 641 false image-claim pairs.

\subsection*{Tools}
In addition to methods, tools to verify the visual content of fake news online is valuable due to its convenience to non-technical users. In this subsection, we introduce some publicly available tools for multimedia content verification.

\textbf{Google Reverse Image Search}: A service of searching by an image from Google. The verifiers may upload the image or input the image URL to find similar images as well as the web pages containing them. Other substitutions like Baidu Images\footnote{https://image.baidu.com/}, provide similar service.

\textbf{FotoForensics}\footnote{http://www.fotoforensics.com/}: A website for forensics analysis of JPEG or PNG image, providing information including error level, hidden pixels, metadata and JPEG quality. Over 3.3 million images were analyzed by the service so far.

\textbf{Image Verification Assistant}\footnote{http://reveal-mklab.iti.gr/reveal/}: A website to analyze the veracity of online media supported by REVEAL project. For an image, it extracts and visualizes the metadata and detects various types of forensics features, such as double JPEG quantization, JPEG Ghosts, JPEG blocking artifact, error level analysis, high-frequency noise and median filtering noise residue.

\textbf{Fake Video News Debunker}\footnote{https://www.invid-project.eu/tools-and-services/invid-verification-plugin/}: A free plugin that runs in Google Chrome or FireFox to verify videos and images. This integrated plugin provides service to obtain contextual information from Youtube or Facebook, extract keyframes for reverse image search, list the metadata and perform forensic analysis.

\subsection*{Relevant Competitions}
To attract the attention from academia and industry and further promote the development of detection technology, considerable competitions for fake news detection were held but very few of them provided visual contents. Here, we introduce two competitions where visual contents can be exploited.

\textbf{Verifying Multimedia Use (VMU)}: A part of the MediaEval Benchmark in 2015\cite{mediaeval15} and 2016\cite{mediaeval16}, dealing with the automatic detection of manipulation and misuse of web multimedia content. A fake tweet was defined as a tweet that shared multimedia content inconsistent with the event it referred to. In 2015, participants were asked to predict the veracity (fake, real or unknown), given a tweet and the accompanying multimedia item (image or video) from an event. In 2016, a new related sub-task was added to detect image tampering.

\textbf{False News Detection Competition 2019}\footnote{https://www.biendata.com/competition/falsenews/}: A competition held for false news detection on Weibo, with three sub-tasks: text-only, image-only and multi-modal detection. In image-only detection, models had to predict whether the image was attached to a false news post. In multi-modal detection, text, images and user profiles were all available to predict the veracity of the post.

\end{document}